\documentclass{elsart}

\usepackage{graphicx}
\usepackage{amssymb}
\usepackage{amsxtra,amssymb,amsmath,latexsym}

\theoremstyle{plain}

\newtheorem{lemma}{Lemma}

\def\nd{\noindent}

\def\R{{\mathbb R}}

\def\oH{\buildrel\circ\over H}
\def\oH1{\buildrel\circ\over H\kern-.02in{}^1}

\def\ve{{\varepsilon}}

\def\tildeD{\widetilde D}

\def\bysame{\rule{.5in}{.005in},\ }

\def\bee{\begin{equation*}}
\def\eee{\end{equation*}}
\def\be{\begin{equation}}
\def\ee{\end{equation}}

\begin{document}

\begin{frontmatter}
\title{Distribution of particles which produces a desired radiation 
pattern}
\author{\underline{Alexander G.~Ramm}}
\address{Mathematics Department, Kansas State University,
 Manhattan, KS 66506-2602, USA\\ phone: 785-532-0580, \quad fax: 785-532-0546 }
\ead{ramm@math.ksu.edu}

\begin{abstract}%
If $A_q(\beta, \alpha, k)$ is the scattering amplitude, corresponding to
a potential $q\in L^2(D)$, where $D\subset\R^3$ is a bounded domain,
and $e^{ik\alpha \cdot x}$ is the incident plane wave, then
we call the radiation pattern the function $A(\beta):=A_q(\beta, \alpha,
k)$, where the unit vector $\alpha$,
the incident direction, is fixed, and $k>0$, the wavenumber, is fixed.
It is shown that any function $f(\beta)\in L^2(S^2)$,
where $S^2$ is the unit sphere in $\R^3$, can be
approximated with any desired accuracy by a radiation pattern:
$||f(\beta)-A(\beta)||_{L^2(S^2)}<\epsilon$, where $\epsilon>0$
is an arbitrary small fixed number. The potential $q$, corresponding to
$A(\beta)$, depends on $f$ and $\epsilon$.

There is a one-to-one correspondence between the above potential and
the density of the number of small
acoustically soft particles $D_m\subset
D$, $1\leq m\leq M$, distributed in an a priori given bounded domain
$D\subset\R^3$.  The geometrical shape of a small
particle $D_m$ is arbitrary, the boundary $S_m$ of $D_m$ is Lipschitz
uniformly with respect to $m$. The wave number $k$ and the direction
$\alpha$ of the incident upon $D$ plane wave are fixed.

It is shown that a suitable distribution of the above particles in
$D$ can produce the scattering amplitude $A(\alpha',\alpha)$,
$\alpha',\alpha\in S^2$, at a fixed $k>0$, arbitrarily close in the
norm of $L^2(S^2\times S^2)$ to an arbitrary given scattering amplitude
$f(\alpha',\alpha)$, corresponding to a real-valued potential $q\in
L^2(D)$, i.e., corresponding to an arbitrary given refraction coefficient in 
$D$. 
\end{abstract}

\end{frontmatter}

\section{Introduction}\label{S1}

It is proved that one can distribute in a given bounded region $D$, small 
in comparison with the wavelength in $D$,  
acoustically soft particles, so that the resulting 
``smart" material will have the scattering 
amplitude $A(\beta,\alpha)$ at a \textit{fixed} wavenumber $k>0$, which 
approximates 
an arbitrary given scattering amplitude, corresponding to any given 
refraction coefficient, with any desired accuracy.

We denote by $\lambda$ the wavelength in $D$, by $a$ the characteristic 
dimension of small particles. We assume that  $a\ll \lambda$.
By $k>0$ we denote the wavenumber in the free space, outside $D$. 

It is also  proved that one can distribute small particles in $D$
so that the corresponding radiation pattern, i.e., the scattering amplitude  
$A(\beta):A(\beta, \alpha)$ at a \textit{fixed} $k>0$ and a \textit{ fixed} 
$\alpha\in S^2$,
would approximate with any desired accuracy an arbitrary given function 
$f(\beta)\in L^2(S^2)$. 
Here 
$S^2$ is the unit 
sphere in $\R^3$, $\alpha$ is the direction of the incident plane wave 
$u_0:=e^{ik\alpha\cdot x}$, $r=|x|$, and  $\beta:=\frac{x}{r}$ is 
a 
direction of the scattered wave. Let us formulate the statement of the 
problem and the results. 

The governing equation is
\be\label{e1}
 [\nabla^2+k^2n_0(x)]u:=[\nabla^2+k^2-q_0(x)]u=0
 \hbox{\ in\ } \R^3, \qquad k=const>0,\ee
$n_0(x)$ is a given refraction coefficient, $n_0(x)=1$ in 
$D'=\R^3\backslash D$, $D$ is a bounded domain, $n_0(x)>0$ in $\R^3$, 
$u$ is acoustic pressure, 
\[q_0(x) :=k^2[1-n_0(x)]=0 \hbox{\quad  in \quad} D'.\]
The scattering problem consists of finding the scattering solution $u$ to 
\eqref{e1} with the asymptotics
\be\label{e2}
  u=u_0+A_0(\beta,\alpha) \frac{e^{ikr}}{r}+o\left(\frac{1}{r}\right), 
  \quad r:=|x|\to\infty,\ \beta:=\frac{x}{r},\ u_0:=e^{ik\alpha\cdot x}.\ee
The coefficient $A_0(\beta,\alpha)$ is called the scattering amplitude, 
 corresponding to $q_0$, or  the radiation 
pattern. If many small particles 
$D_m$, $1\leq m\leq M$, are embedded in $D$, $u\mid_{S_m}=0$, where $S_m$ 
is the boundary of $D_m$, then
the scattering problem is:
\be\label{e3}
 [\nabla^2+k^2-q_0(x)]u=0\hbox{\ in\ }\R^3\setminus \bigcup^M_{m=1}D_m \ee
\be\label{e4}  u\mid_{S_m}=0, \quad m=1,\dots,M,\ee 
\be\label{e5}
 u=u_0+A(\beta,\alpha)\,\frac{e^{ikr}}{r}+o\left(\frac{1}{r}\right),
 \quad r=|x|\to\infty,\quad \beta=\frac{x}{r},\ee
and the solution $u$ is called the scattering solution.

We assume
\be\label{e6}  ka\ll 1,\qquad a\ll d, \ee
where \[a=\frac{1}{2} \max_{1\leq m\leq M} diam\,D_m,\quad 
d=\min_{m\not= j}\,dist\,(D_m,D_j).\] Assumptions \eqref{e6} allow us 
to have $d\ll \lambda$. We assume that the quantity 
$\frac{a}{d^3}$ has a finite non-zero limit as $M\to\infty$ and 
$\frac{a}{d}\to0$. 
More precisely, if $C_m$ is the electrical capacitance of the conductor 
with the shape $D_m$, then we assume the existence of a limiting density 
of the capacitance per unit volume around every point $x\in D$:
\be\label{e7}
 \lim_{M\to\infty}\sum_{D_m\subset\tildeD} C_m=\int_{\tildeD} C(x)dx, \ee
where $\tildeD\subset D$ is an arbitrary subdomain of $D$. Note that the 
density of the volume of the small particles per unit volume is 
$O\left(\frac{a^3}{d^3}\right)\to 0$ as $\frac{a}{d}\to 0$.
One can prove (see \cite{R476}, p.103) that in the limit $M\to\infty$ the 
function $u$ solves the equation
\be\label{e8}
 [\nabla^2+k^2-q(x)]u=0 \hbox{\ in\ }\R^3,\quad q(x)=q_0(x)+C(x),\ee
where $C(x)$ is defined in \eqref{e7}, and $A(\beta,\alpha)$ in \eqref{e5} 
corresponds to the potential $q(x)$   (see also \cite{MaKh}, where similar 
homogenization-type problems are discussed).

In \cite{R470} the problem of finding $q$ from the data $A(\beta,\alpha)$ 
known for all $\beta,\alpha\in S^2$ (actually, for all $\beta\in S^2_1$ 
and all $\alpha\in S^2_2$, where $S^2_j$, $j=1,2$, are arbitrary open 
subsets in $S^2$) has been investigated in detail: uniqueness of its 
solution is 
proved, an algorithm for recovery of $q$ from the above data was 
proposed, and its error estimate was obtained in the cases of both
exact and noisy 
data.

If all the small particles are identical, $C_0$ is the capacitance of a 
conductor in the shape of a particle, and $N(x)$ is the number of small 
particles per unit volume around point $x$, then \[C(x)=N(x)C_0.\] 
Therefore, 
\[q(x)=q_0(x)+N(x)C_0,\] 
and
\be\label{e9}  N(x)=\frac{q(x)-q_0(x)}{C_0}. \ee
Thus, one has an explicit one-to-one correspondence between $q(x)$ and the 
density $N(x)$ of the embedded particles per unit volume. 
Let us describe our solution to problem $(P_1)$, which consists of finding 
$N(x)$, given $A(\beta,\alpha)$.

If $A(\beta,\alpha)$ is the scattering amplitude, corresponding to a 
potential $q\in L^2(D)$, then, given $A(\beta,\alpha)$, one applies 
Ramm's algorithm (see \cite{R470}, Chapter 5), recover $q(x)$, and 
then calculate $N(x)$ by 
formula \eqref{e9}.

Consider a new inverse problem ($P_2$): 

{\it Let $A(\beta)\in L^2(S^2)$ be 
an arbitrary funceion. Can one find a potential $Q\in L^2(D)$, such 
that the corresponding scattering amplitude $A_q(\beta)$ at a fixed $k>0$ 
and a fixed $\alpha\in S^2$ would approximate $f(\beta)$ with a desired 
accuracy:}
\be\label{e10}  \|f(\beta)-A_q(\beta)\|<\ve. \ee

The answer is yes. Let us explain this. Note that
\be\label{e11}
 A_q(\beta)=-\frac{1}{4\pi}\int_D e^{-ik\beta\cdot x}q(x)u(x)dx,\ee
where $u(x)$ is the scattering solution corresponding to a potential $q\in 
L^2(D)$, $k$ and $\alpha$ are fixed. If $q$ is a known, 
then $N(x)$ can be calculated by \eqref{e9}. Denote
\be\label{e12}   h:=q(x)u(x). \ee

\begin{lemma}\label{lemma1}
For any $f\in L^2(S^2)$ and any $\ve_1>0$, however small, there exists an 
$h=h_{\ve_1}(x)\in L^2(D)$, such that
\be\label{e13}
 \|f(\beta)+\frac{1}{4\pi} \int_D e^{-ik\beta\cdot x} 
 h(x)dx\|_{L^2(S^2)} <\ve_1. \ee
\end{lemma}

Thus, given $f$ and $\ve_1>0$, one can find $h\in L^2(D)$ such that 
\eqref{e13} holds. This $h$ can be found either numerically or 
analytically 
(see \cite{R515}).

If $h$ is found, then one can find a $q(x) \in L^2(D)$ such that
\be\label{e14}  \|h-q(x)u(x)\|_{L^2(D)} <\ve_2, \ee
where $\ve_2>0$ is an arbitrary small given number. 

This is possible due to the following result.

\begin{lemma}\label{lemma2}
For any $h\in L^2(D)$ and any $\ve_2>0$, there exists a $q\in L^2(D)$, 
such that \eqref{e14} holds.
\end{lemma}

If such a $q$ is found, then \eqref{e10} holds with an arbitrary small 
$\ve>0$, if $\ve_1$ and $\ve_2$ are sufficiently small. This follows from 
Lemmas 1, 2 and formula \eqref{e11}.

Let us describe the steps of the solution to Problem $(P_2)$.

\nd\underbar{Step 1.} Given an arbitrary function $f(\beta)\in L^2(S^2)$ 
and $\ve_1>0$, one finds $h\in L^2(D)$ such that \eqref{e13} holds.

\nd\underbar{Step 2.} Given $h$ and $\ve_2>0$, one finds $q\in L^2(D)$ 
such that \eqref{e14} holds.

\nd\underbar{Step 3.}
This $q$ generates the scattering amplitude $A(\beta)$ at fixed $\alpha$ 
and $k$, such that \eqref{e10} holds.

Indeed, let $\|\cdot\|:=\|\cdot\|_{L^2(S^2)}$. Then
\be\label{e15}\begin{aligned}
 \|f(\beta)-A_q(\beta)\| 
 &=\|f(\beta)+\frac{1}{4\pi}\int_D e^{-ik\beta\cdot x} q(x)u(x)dx\| \\
 &\leq\|f(\beta) +\frac{1}{4\pi}\int_D e^{ik\beta\cdot x}hdx\| \\
 &+\frac{\ve_2|D|}{4\pi}\leq \ve_1+\frac{\ve_2|D|}{4\pi}<\ve,
 \quad |D|=meas\,D. \end{aligned}\ee

{\bf Remark.} If the boundary condition on $S_m$ is of impedance type:
\[  u_N=\zeta u\hbox{\quad on\quad} S_m, \]
where $N$ is the exterior unit normal to the boundary $S_m$, and $\zeta$ is
a complex constant, the impedance, then the capacitance $C_0$ 
in formula (9) should be replaced by
\[ C_\zeta =\frac{C_0}{1+\frac{C_0}{ \zeta |S|}},\]
where $|S|$ is the surface area of $S$, and the corresponding potential 
$q(x)$ will be complex-valued 
(see \cite{R476}, p. 97).

\end{document}